\documentclass{desyproc}

\begin{document}
\title{Dilepton Production in Transport-based Approaches}

\author{{\slshape Janus Weil$^1$, Stephan Endres$^1$, Hendrik van Hees$^1$, 
Marcus Bleicher$^1$, Ulrich Mosel$^2$}\\[1ex]
$^1$Frankfurt Institute for Advanced Studies , Ruth-Moufang-Str.~1, 60438 
Frankfurt, Germany\\
$^2$Institut f\"ur Theoretische Physik, JLU Giessen, Heinrich-Buff-Ring 16, 
35392 Giessen, Germany}

\contribID{173}

\confID{8648}  
\desyproc{DESY-PROC-2014-04}
\acronym{PANIC14} 
\doi  

\maketitle

\begin{abstract}
We investigate dilepton production in transport-based approaches and show that 
the baryon couplings of the $\rho$ meson represent the most important 
ingredient for understanding the measured dilepton spectra. At low energies (of 
a few GeV), the baryon resonances naturally play a larger role and affect 
already the vacuum spectra via Dalitz-like contributions, which can be captured 
well in an on-shell-transport scheme. At higher energies, the baryons mostly 
affect the in-medium self energy of the $\rho$, which is harder to tackle in 
transport models and requires advanced techniques.
\end{abstract}


\section{Introduction}

Lepton pairs are known to be an ideal probe for studying phenomena at high 
densities and temperatures. They are created at all stages of a heavy-ion 
collision, but unlike hadrons they can escape the hot and dense zone almost 
undisturbed (since they only interact electromagnetically) and thus can carry 
genuine in-medium information out to the detector. Dileptons are particularly 
well-suited to study the in-medium properties of vector mesons, since the 
latter can directly convert into a virtual photon, and thus a lepton pair 
\cite{Leupold:2009kz,Rapp:2009yu}.
One of the groundbreaking experiments in this field was NA60 at the CERN SPS, 
which revealed that the $\rho$ spectral function is strongly broadened in the 
medium. Calculations by Rapp et al. have shown that this collisional broadening 
is mostly driven by baryonic effects, i.e., the coupling of the $\rho$ meson to 
baryon resonances ($N^*$, $\Delta^*$) \cite{vanHees:2006ng}.
In the low-energy regime, the data taken by the DLS detector have puzzled 
theorists for years and have recently been confirmed and extended by new 
measurements by the HADES collaboration 
\cite{Agakishiev:2009yf,HADES:2011ab,Agakishiev:2012tc,Agakichiev:2006tg, 
Agakishiev:2007ts,Agakishiev:2011vf}. At such low energies, it is expected that 
not only the in-medium properties are determined by baryonic effects, but that 
already the production mechanism of vector mesons is dominated by the coupling 
to baryons (even in vacuum).


\section{The model: hadronic transport + VMD}

\label{sec:VMD}

\begin{figure}[t]
  \hspace{0.05\textwidth}
  \includegraphics[width=0.35\textwidth]{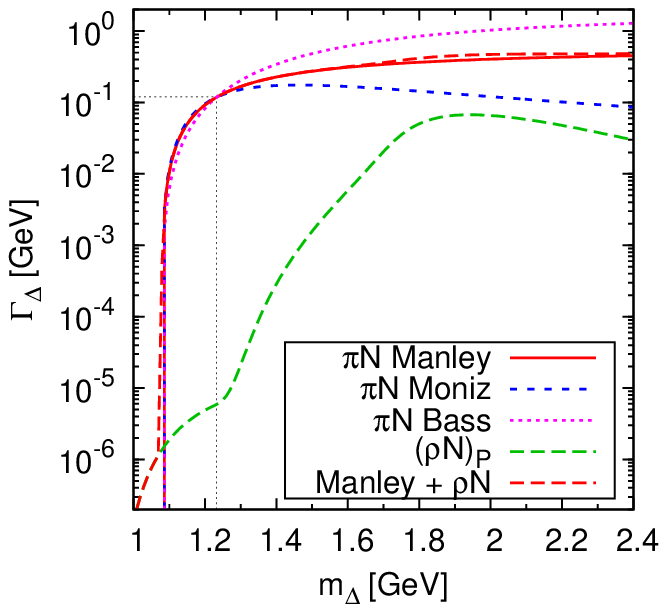}
  \hspace{0.1\textwidth}
  \includegraphics[width=0.35\textwidth]{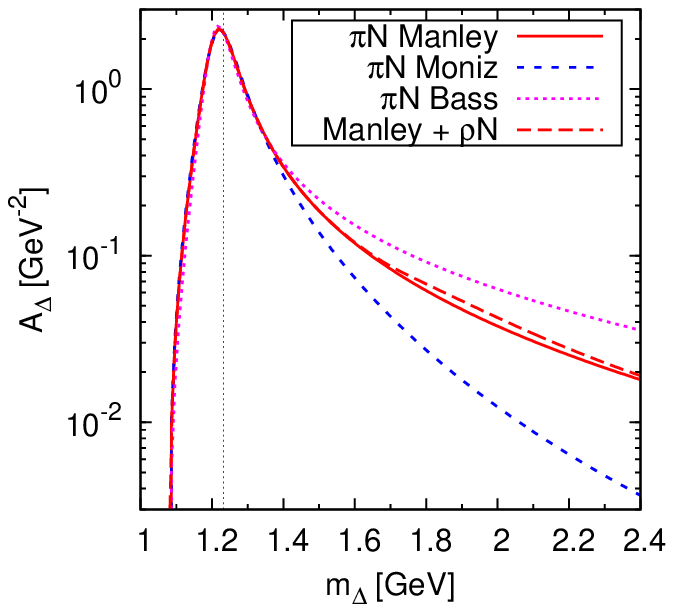}
  \hspace{0.05\textwidth}
  \caption{Partial widths for the $\pi N$ and $\rho $N decay channels (left) 
and spectral function (right) of the $\Delta$ resonance as a function of the 
off-shell mass.}
  \label{fig:Delta_rhoN}
\end{figure}

Already our previous investigations \cite{Weil:2012ji} based on the GiBUU 
transport model \cite{Buss:2011mx} have shown that the baryonic $N^*$ and 
$\Delta^*$ resonances can give important contributions to dilepton spectra at 
SIS energies, both from pp and AA collisions, via Dalitz-like contributions.
This finding was based on the assumption that these resonances decay into a 
lepton pair exclusively via an intermediate $\rho$ meson (i.e. strict 
vector-meson dominance).
In the transport simulation, the Dalitz decays $R\rightarrow e^+e^-N$ are 
treated as a two-step process, where the first part is an $R\rightarrow\rho N$ 
decay, followed by a subsequent conversion of the $\rho$ into a lepton pair 
($\rho\rightarrow e^+e^-$).
The branching ratios for the $R\rightarrow\rho N$ decay are taken from the 
partial-wave analysis by Manley et al.~\cite{Manley:1992yb},
while the decay width for the second part is calculated under the strict-VMD 
assumption as $\Gamma_{ee}(m)=\Gamma_0\cdot(M/m)^3$.
For the present study we extend the VMD assumption also to the $\Delta(1232)$ 
state, whose dilepton contribution has been subject to much controversy 
recently. Since the $\Delta$ is too light to decay into an on-shell $\rho$ 
meson, it is difficult to determine its coupling to the $\rho$ experimentally, 
and consequently Manley and other analyses do not find any sign of a 
$\Delta\rightarrow\rho N$ decay. Nevertheless the $\Delta$ has a photonic decay 
mode, which means that also a dilepton Dalitz decay channel must exist. The 
latter has been claimed to be particularly significant for dilepton spectra at 
SIS energies \cite{Bratkovskaya:2013vx}. However, this argument was based on 
the continuation of the photon decay into the time-like region neglecting the 
involved electromagnetic transition form factor \cite{Krivoruchenko:2001hs}. 
Unfortunately this form factor is essentially unknown in the time-like region 
from the experimental point of view. However, it is clear that it can 
significantly alter the dilepton yield from the $\Delta$ (easily by an order of 
magnitude) \cite{Ramalho:2012ng}.
In order to deal with this situation, we choose to apply the assumption of 
strict VMD not only to the $N^*$ and $\Delta^*$ resonances, but also to the 
$\Delta$ itself, assuming a p-wave (i.e.~$L=1$) decay into $\rho N$. Together 
with the other resonance channels, this results in a consistent model with 
clear assumptions, which can be tested against experiment.
One free parameter that is left to fix in this approach is the on-shell 
branching ratio of $\Delta\rightarrow\rho N$. We use a value of 
$5\cdot10^{-5}$, in order to produce dilepton yields which are roughly 
equivalent to the radiative decay for small $\Delta$ masses and compatible with 
the HADES data at low energies. Fig.~1 shows the partial decay width into $\rho 
N$, which is extremely small at the $\Delta$ pole mass, but grows significantly 
when going to larger masses. But even in the very high-mass tail, the 
additional decay mode has only little influence on the overall width and 
spectral function of the $\Delta$ (even less than the different parametrizations 
of the $\pi N$ width).


\section{Dilepton spectra from p+p collisions}

Fig.~2 shows a comparison of our simulation results for p+p collisions (inside 
the detector acceptance) to the dilepton mass spectra measured by the HADES 
collaboration at three different beam energies.
Since the mesonic decay channels have not changed with respect to earlier works 
\cite{Weil:2012ji}, we concentrate here on the discussion of the baryonic 
contributions. The $\Delta$ is shown in two approaches, a QED-like radiative 
decay \cite{Krivoruchenko:2001hs} (neglecting the occurring form factor), and a 
VMD decay $\Delta\rightarrow\rho N\rightarrow e^+e^-N$. While both give rather 
similar results at low energies (where the form-factor effects are still 
small), the differences get larger at higher energies. There the VMD curve 
develops a clear peak at the $\rho$ mass and a bump around 
$m_{\Delta}-m_N\approx300$ MeV (from the on-shell $\Delta$s), while the QED 
curve is flat and structureless (due to the absence of a form factor). However, 
both models agree on the fact that the $\Delta$ contribution becomes 
sub-dominant at higher energies and is exceeded by other contributions (in 
particular the higher resonances $N^*$ and $\Delta^*$ become more significant). 
Thus the data can not distinguish between both models.

\begin{figure}[ht]
  \includegraphics[width=\textwidth]{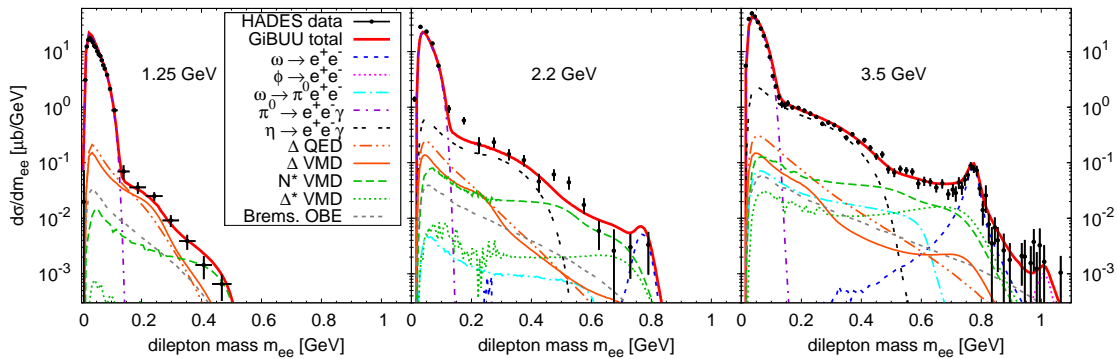}
  \caption{Dilepton mass spectra for pp collisions, in comparison to the data 
from \cite{Agakishiev:2009yf,HADES:2011ab,Agakishiev:2012tc}.}
  \label{fig:pp}
\end{figure}


\section{Dilepton spectra from A+A collisions}

Fig.~3 shows our results of dilepton spectra from nucleus-nucleus collisions 
compared to the HADES data. The light CC system has been measured at two 
different energies (1 and 2 AGeV) and the heavier ArKCl at the intermediate 
energy of 1.76 AGeV. The best agreement with data is achieved in the CC system 
at 2 AGeV, where the spectrum above the pion mass is dominated by the $\eta$ 
Dalitz and the baryonic VMD channels. In the 1 AGeV reaction, we see some 
underestimation at intermediate masses around 300 MeV, despite the inclusion of 
OBE Bremsstrahlung according to Shyam et al.~\cite{Shyam:2010vr}. Since there 
are many channels contributing with similar strength here, it is hard to tell 
where the underestimation originates from.
In the medium-size ArKCl system, we see a similar underestimation at 
intermediate masses and a slight excess in the vector-meson pole region. One 
may be surprised that a pure (on-shell) transport approach without explicit 
inclusion of in-medium spectral functions achieves such a good agreement here, 
but that just shows the importance of Dalitz-like contributions of the baryons, 
which are captured well by our transport treatment.

\begin{figure}[ht]
  \includegraphics[width=\textwidth]{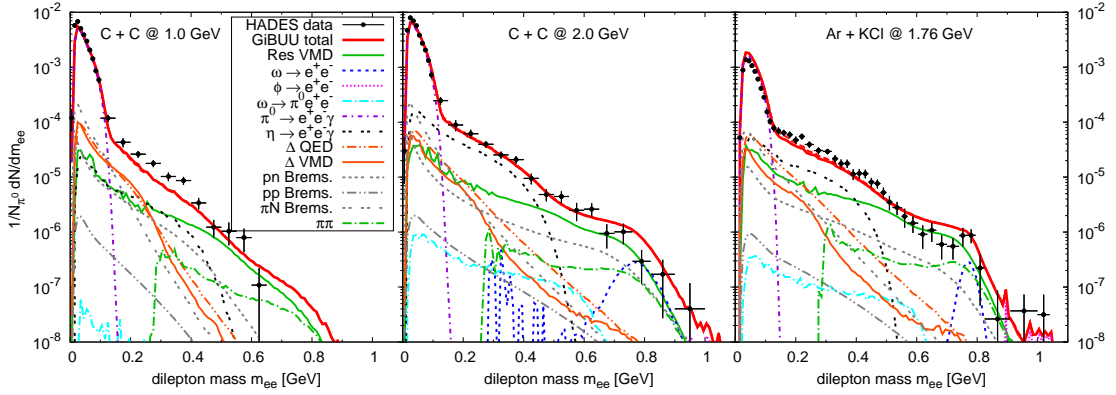}
  \caption{Dilepton mass spectra for AA collisions, in comparison to the data 
from \cite{Agakichiev:2006tg,Agakishiev:2007ts,Agakishiev:2011vf}.}
  \label{fig:AA}
\end{figure}


\section{Conclusions}

We have shown that the HADES dilepton data from pp and AA collisions can be 
described rather well with a combination of a resonance-model-based transport 
approach with a strict-VMD coupling of the baryons to the em. sector, where a 
mix of different baryonic resonances contributes to the total dilepton yield. 
We can not reproduce the dominant contribution of the $\Delta(1232)$, which was 
claimed in other models \cite{Bratkovskaya:2013vx}.
In order to improve the description of heavy systems and to make the connection 
to higher energies, a proper dynamic treatment of in-medium spectral functions 
is required, which may be provided by the so-called ``coarse-graining`` 
approach, which is subject of ongoing investigations \cite{Endres:2013cza}.

\section*{Acknowledgments}

This work was supported by the Hessian Initiative for Excellence (LOEWE) 
through the Helmholtz International Center for FAIR and by the Federal Ministry 
of Education and Research (BMBF). J.W.~acknowledges funding of a Helmholtz 
Young Investigator Group VH-NG-822 from the Helmholtz Association and GSI.


\begin{footnotesize}

\end{footnotesize}


\end{document}